\documentclass[preprint,review,12pt]{elsarticle}

\usepackage{graphicx}
\usepackage{amssymb}

\journal{Coastal Engineering}

\begin{document}

\begin{frontmatter}


\title{A Machine Learning Framework \\ to Forecast Wave Conditions}


\author[1]{Scott C. James}
\author[2]{Yushan Zhang}
\author[3]{Fearghal O'Donncha}
\address[1]{Baylor University, Deptartments of Geosciences and Mechanical Engineering, One~Bear~Place~\#97354, Waco, TX  76798-2534.}
\address[2]{University of Notre Dame, Department of Chemical and Biomolecular Engineering, Notre Dame, IN 46556-5637.}
\address[3]{IBM Research, Dublin, Ireland.}

\begin{abstract}
A~machine learning framework is developed to estimate ocean-wave conditions. By supervised training of machine learning models on many thousands of iterations of a physics-based wave model, accurate representations of significant wave heights and period can be used to predict ocean conditions. A~model of Monterey Bay was used as the example test site; it was forced by measured wave conditions, ocean-current nowcasts, and reported winds. These input data along with model outputs of spatially variable wave heights and characteristic period were aggregated into supervised learning training and test data sets, which were supplied to machine learning models. These machine learning models replicated wave heights with a root-mean-squared error of 9~cm and correctly identify over 90\% of the characteristic periods for the test-data sets. Impressively, transforming model inputs to outputs through matrix operations requires only a fraction ($<1/1,000^\mathrm{th}$) of the computation time compared to forecasting with the physics-based model.
\end{abstract}


\end{frontmatter}


\section{Introduction}
\label{S:1}
There are myriad reasons why predicting wave conditions is important to the economy. Surfers aside, there are fundamental reasons why knowledge of wave conditions for the next couple of days is important. For example, shipping routes can be optimized by avoiding rough seas thereby reducing shipping times. Another industry that benefits from knowledge of wave conditions is the \$160B (2014) aquaculture industry \citep{fao2016}, which could optimize harvesting operations accordingly. Knowledge of littoral conditions is critical to military and amphibious operations by Navy and Marine Corps teams. Also, predicting the energy production from renewable energy sources is critical to maintaining a stable electrical grid because many renewable energy sources (e.g.,~solar, wind, tidal, wave, etc.) are intermittent. For deeper market penetration of renewable energies, combinations of increased energy storage and improved energy-generation predictions will be required. The US~Department of Energy has recently invested in the design, permitting, and construction of an open-water, grid-connected national Wave Energy Test Facility at Oregon State University \citep{USDOE_2016}. Given that America's technically recoverable wave-energy resource is up to 1,230~TW-hr \citep{EPRI_2011}, there is a strong interest in developing this renewable resource \citep{OceanEnergySystems_2016}. Commercialization and deployment of wave-energy technologies will require not only addressing permitting and regulatory matters, but overcoming technological challenges, one of which is being able to provide an accurate prediction of energy generation. A~requirement for any forecast is that an appropriately representative  model be developed, calibrated, and validated. Moreover, this model must be able to run extremely fast and to incorporate relevant forecast data into its predictions. A~machine learning framework for this capability is developed here.

Because wave models can be computationally expensive, a new approach with machine learning \citep{goodfellow2016deep,lecun2015deep,schmidhuber2015deep} is developed here. The goal of this approach is to train machine learning models on many realizations of a physics-based wave model forced by historical atmospheric and sea states to accurately represent wave conditions (specifically, significant wave heights and characteristic period). Predicting these wave conditions at locations corresponding to a (potential) wave-energy-converter (WEC) array facilitates accurate power-production forecasts. Given the recent development of a wave-energy-resource classification system \citep{haas_et_al_2017}, if the waves conditions at a particular location can be predicted, the power potential for a hypothetical WEC array can be estimated.

Computational expense is often a major limitation of real-time forecasting systems \citep{devries2017enabling,mallet2009ozone}. Here, we apply machine learning techniques to predict wave conditions with the goal of replacing a computationally intensive physics-based model by straightforward multiplication of an input vector by mapping matrices resulting from the trained machine learning models. Because matrix multiplication is an exceedingly rapid operation, the end result is a machine-learning technique that can predict wave conditions with comparable accuracy to a physics-based model for a fraction of the computational cost. While machine learning has been used to predict wave conditions \cite{PERES2015128,makarynskyy2004improving,ETEMADSHAHIDI20091175,mahjoobi2008alternative,1715295,BROWNE2007445}, it has not been used in the context of a surrogate model as defined below.

One of the challenges for machine learning applications is their enormous appetite for data. It is the exception more than the rule that a machine learning approach has what is considered an optimal amount of data available to it. However, when developing a machine learning surrogate for a physics-based model, there is the luxury of being able to run the model as many times as necessary to develop a sufficient data set to train the machine learning model. Here, we define a surrogate model \citep{razavi2012review} as a data-driven technique to empirically approximate the response surface of a physics-based model. These have alternately been called ``metamodels'' \citep{blanning-1975,KLEIJNEN2009707},``model emulators'' \citep{OHAGAN20061290}, and ``proxy models'' \citep{bieker2007real}.

\section{Wave Modeling}
\subsection{Numerical Model}
The Simulating WAves Nearshore (SWAN) FORTRAN code is the industry-standard wave-modeling tool developed at the Delft University of Technology that computes wave fields in coastal waters forced by wave conditions on the domain boundaries, ocean currents, and winds \citep{swan_tech_man}. SWAN models the energy contained in waves as they travel over the ocean and disperse at the shore. Specifically, information about the sea surface is contained in the wave-variance spectrum, or energy density $E\left(\sigma,\theta\right)$, and this wave energy is distributed over wave frequencies (as observed in an inertial frame of reference moving with the current velocity) with propagation directions  normal to wave crests of each spectral component. 

Action density is defined as $N = E/\sigma$, which is conserved during propagation along the  wave characteristic in the presence of ambient current. Evolution of $N\left(x,y,t;\sigma,\theta\right)$ in space, $x,y$, and time, $t$, is governed by the action balance equation \citep{komen1996dynamics,mei1989theory}:
\begin{equation}
\frac{\partial N}{\partial t} + \left( \frac{\partial c_x N}{\partial x} + \frac{\partial c_y N}{\partial y}\right) + \left( \frac{\partial c_\sigma N}{\partial \sigma} + \frac{\partial c_\theta N}{\partial \theta}\right) = \frac{S_{\mathrm{tot}}}{\sigma}.
\label{eqn:wave}
\end{equation}
The left-hand side represents the kinematic component of the equation. The second term (parenthetical) denotes the propagation of wave energy in a two-dimensional Cartesian space where $c$ is wave celerity. The third term represents the effect of shifting of the radian frequency due to variations in water depth and mean current. The fourth term expresses depth- and current-induced refractions. The quantities $c_\sigma$ and $c_\theta$ are the propagation speeds in spectral space $(\sigma,\theta)$. The right-hand side represents the spatio-temporally variable sources and sinks of all physical processes that generate, dissipate, or redistribute wave energy (i.e.,~wave growth by wind, nonlinear transfer of energy through three- and four-wave interactions, and wave decay due to white-capping, bottom friction, and depth-induced wave breaking).

\citet{haas_et_al_2017} define the wave-energy resource as a function of the significant wave height, $H_\mathrm{s}$ and  peak wave period, $T$. This information can be used to compute the wave power density. Hence, estimates of peak period $T$ and, in  particular, $H_\mathrm{s}$ because $J$ is quadratically related to wave height, are necessary to predict wave energy potential.

\subsection{Model Verification}
The coastal ocean presents a complex modeling challenge, intimately connected as it is to both the deep ocean and the atmosphere \citep{song1994semi}. Uncertainties in wave forecasting emanate from the mathematical representation of the system, numerical approximations, and uncertain and incomplete data sets. Studies demonstrate that the greatest sources of uncertainty in operational wave forecasting are the model input data. This study simulates wave conditions subject to real forcing conditions at a case-study site, Monterey Bay, California. As summarized in Table~\ref{table:data}, the wave model was driven by available NOAA wave-condition data, archived current nowcasts from the Central and Northern California Ocean Observing System (CeNCOOS) \citep{patterson2012addressing}, and post-processed (i.e.,~data subject to quality assurance procedures) wind data from \citet{TWC}. 

\begin{table}\footnotesize
\caption{Data sources.}
\label{table:data}
    \begin{tabular}{| c | c | c | c |}
    \hline
    Data & Source & URL & Resolution \\ \hline
    Wave conditions       & NDBC$^1$ &  \multicolumn{1}{|l|}{http://www.ndbc.noaa.gov/station\_page.php?station=46042} & $36^\circ 47^\prime 29^{\prime\prime}$ N \\
    ($H_\mathrm{s}$, $T$, $D$) & Buoy 46042 & & $122^\circ27^\prime 6^{\prime\prime}$ W \\
    \hline
        Wave conditions       & {\normalsize W}AVE{\normalsize W}ATCH {\normalsize III} & \multicolumn{1}{|l|}{http://nomads.ncep.noaa.gov:9090/dods/wave/enp} & $0.25^\circ$\\
    ($H_\mathrm{s}$, $T$, $D$) &  ENP NCEP$^2$ & &\\
    \hline
    Ocean currents & ROMS COPS$^4$ & \multicolumn{1}{|l|}{http://west.rssoffice.com:8080/thredds/catalog/roms/CA3000m-forecast/catalog.html} & 3 km\\ \hline
    Winds & TWC$^4$ & \multicolumn{1}{|l|}{https://api.weather.com/}  & User defined\\ 
    \hline
    Bathymetry & NOAA NGDC$^5$ &\multicolumn{1}{|l|}{https://www.ngdc.noaa.gov/mgg/bathymetry/hydro.html} & $0.001^\circ$ \\ \hline
    \multicolumn{4}{l} {$^1$National Data Buoy Center} \\
    \multicolumn{4}{l} {$^2$Eastern North Pacific National Centers for Environmental Prediction} \\
    \multicolumn{4}{l} {$^3$Regional Ocean Modeling System Cooperative Ocean Prediction System} \\
    \multicolumn{4}{l} {$^4$The Weather Company}\\
    \multicolumn{4}{l} {$^5$National Geophysical Data Center}
    \end{tabular}
\end{table}

\begin{figure}
\centering
\includegraphics[width=1.0\textwidth]{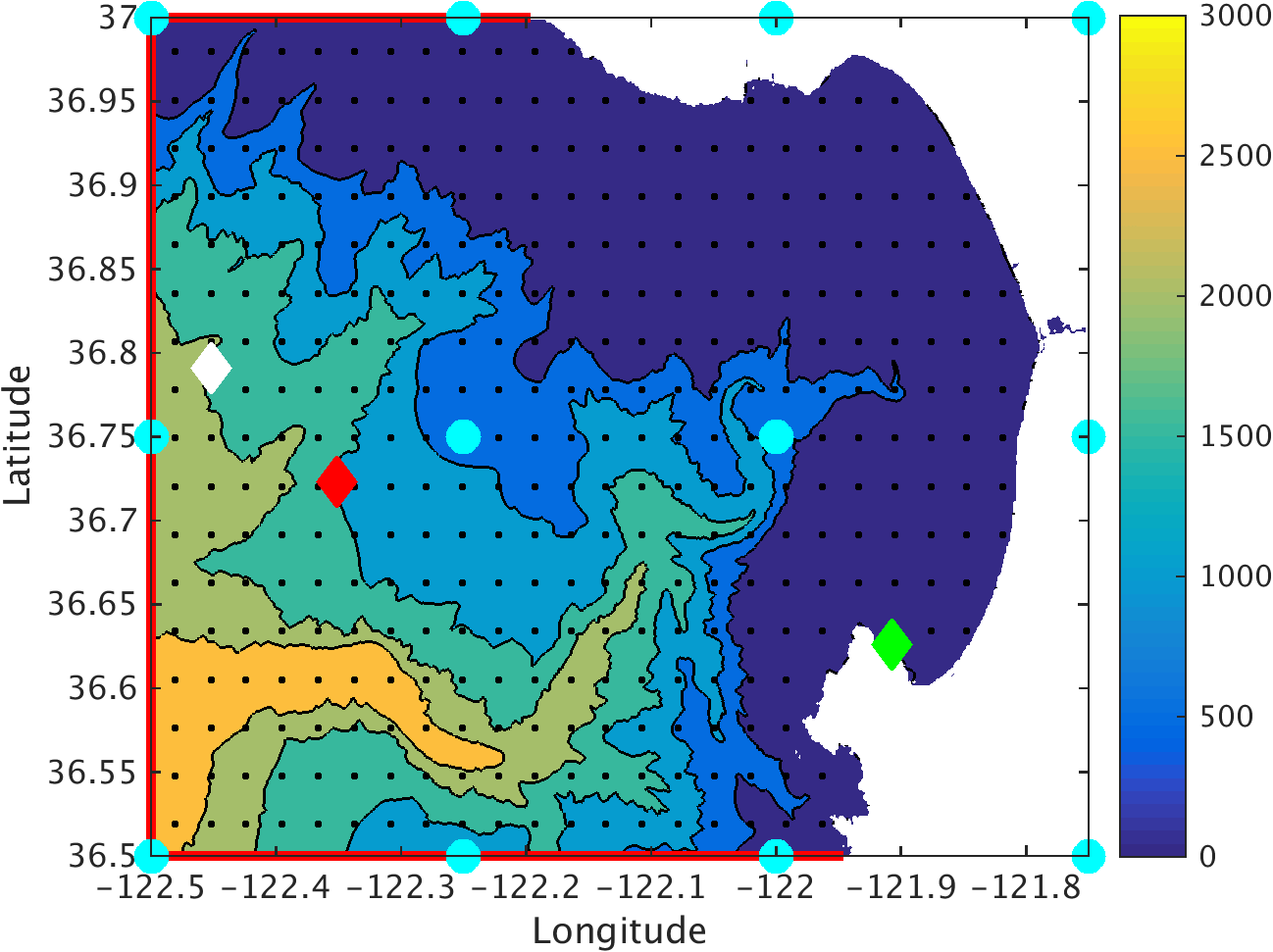}
\caption{SWAN model domain with color indicating the bathymetric depth. The three buoys used to verify the model are indicated with the symbols where the white diamond is Buoy~46042, the red diamond is Buoy~46114, and the green diamond is Buoy~46240.}
\label{fig:bathymetry}
\end{figure}

Before developing a machine-learning surrogate for the physics-based SWAN model, it is important to demonstrate that SWAN can accurately replicate wave conditions in Monterey Bay so that a training data set can be developed for the machine learning models. The SWAN model validated by \citet{chang2016numerical} was used in this effort because it has a demonstrated track record for accurately simulating wave conditions in Monterey Bay. The bathymetric data shown in Figure~\ref{fig:bathymetry} were obtained from the NOAA National Geophysical Data Center. The horizontal resolution in this SWAN model is $0.001^\circ$.   

Because SWAN discretizes wave frequencies in its calculations, only a user-defined number of discrete $T$ values can be returned by a simulation. The user specifies the minimum, $\phi_1$, maximum $\phi_\mathsf{N}$,  and number of discrete frequencies, which are logarithmically distributed as \citep{swan_tech_man}:
\begin{equation}
\phi_\iota = \phi_{\iota-1} + \phi_{\iota-1}\left(\frac{\phi_\mathsf{N}}{\phi_1}-1\right)^\frac{1}{\mathsf{N}-1}.
\end{equation}
Note that the logarithmic distribution yields smaller increments between periods for larger $T$, which is most relevant for capturing the effects of long-period waves -- those most important for energy generation. For these simulations, $\phi_1 = 0.042$~Hz ($T_1=23.8$~s) and $\phi_\mathsf{N} = 1$~Hz ($T_\mathsf{N}=1$~s), and $\mathsf{N}=24$ discrete $T$ values were specified \citep{chang2016numerical}. However, only 11 distinct $T$ values were ever calculated by SWAN because NOAA buoy data ranged from 3.98 to 15.40~s, so henceforth, $\mathsf{N}=11$. Throughout most of the model domain, SWAN calculated a single $T$, with only minimal variation observed in shallow waters comprising about 2\% of the model cells. Hence, a ``characteristic $T$'' is defined for each run of the SWAN model as this is an important wave characteristic for estimating wave energy.

Before developing the machine learning data set, it is important to verify that SWAN can simulate wave characteristics with sufficient accuracy. For this model-verification exercise, inputs forcing the SWAN model comprise wave conditions (defined along the SWAN model boundaries indicated in red in Figure~\ref{fig:bathymetry})  from the NOAA National Data Buoy Center (Buoy~46042), ocean currents from CeNCOOS \citep{patterson2012addressing} (357 $u$ and $v$ currents supplied at the 3-km-spaced small, black squares in Figure~\ref{fig:bathymetry}), and wind data from \citet{TWC} (extracted at 12 locations spaced $0.25^\circ$ apart corresponding to the turquoise circles in Figure~\ref{fig:bathymetry}). The regional wave model is the W{\small AVE}W{\small ATCH}~III \cite{tolman2009user} simulation of the Eastern North Pacific discretized at $0.25^\circ$ (turquoise circles in Figure~\ref{fig:bathymetry}) and its simulations are representative of the typical accuracies in a regional model. Ocean currents are from the Monterey Bay ocean-forecasting system based on the 3-km-resolution Regional Ocean Modeling System (ROMS) \citep{CENCOOS}. ROMS atmospheric forcing is derived from the  Coupled Ocean/Atmosphere Mesoscale Prediction System atmospheric model \citep{hodur1997naval}. Akin to the SWAN model, ROMS wave and tidal forcing are specified at the three lateral boundaries of the outermost domain using information from the global tidal model \citep{dushaw1997topex}. Wind speeds were extracted at $0.25^\circ$ spacing from a TWC application programming interface (API). TWC provides information on a variety of meteorological conditions, forecasts, alerts, and historical data, which can be extracted either directly from the TWC API or through the IBM Bluemix platform. Hourly forecast data out to fifteen days are available along with historical cleansed data for the past 30~years.

\begin{figure}
\centering
\includegraphics[width=1.0\textwidth]{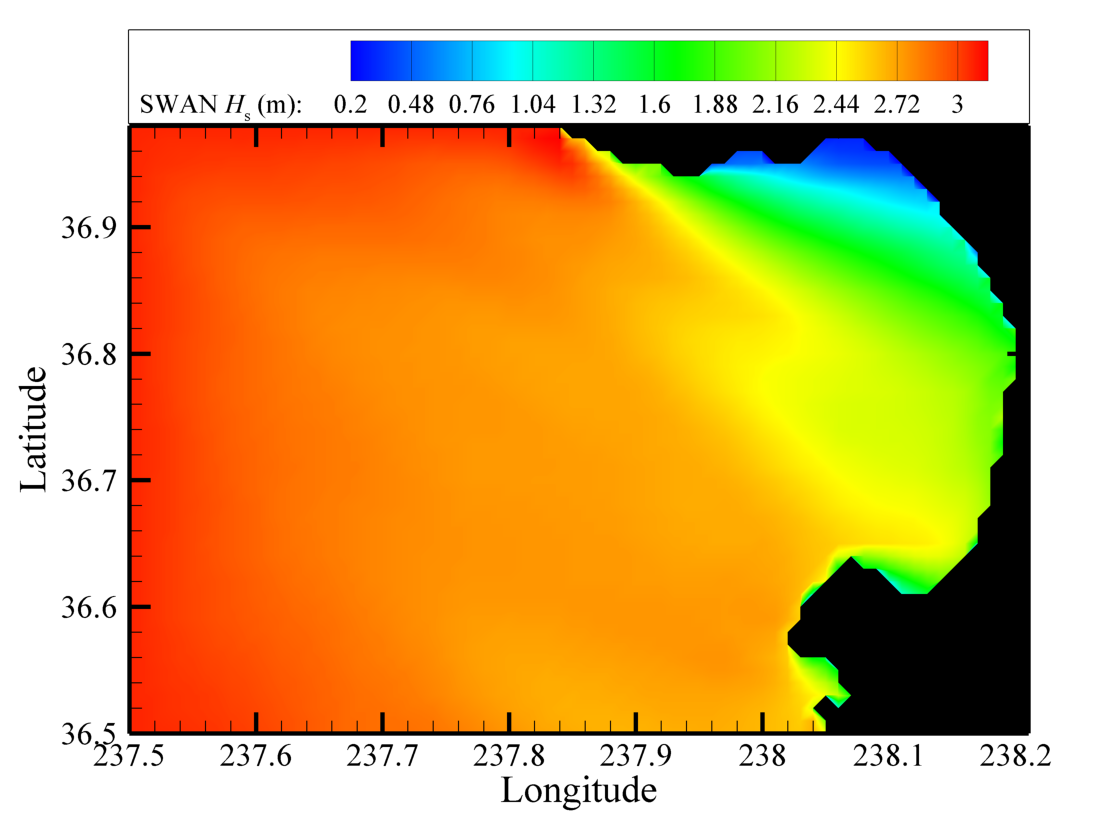}
\caption{Representative SWAN $H_\mathrm{s}$ field.}
\label{fig:SWANHs}
\end{figure}

\begin{figure}
\centering
\includegraphics[width=1.0\textwidth]{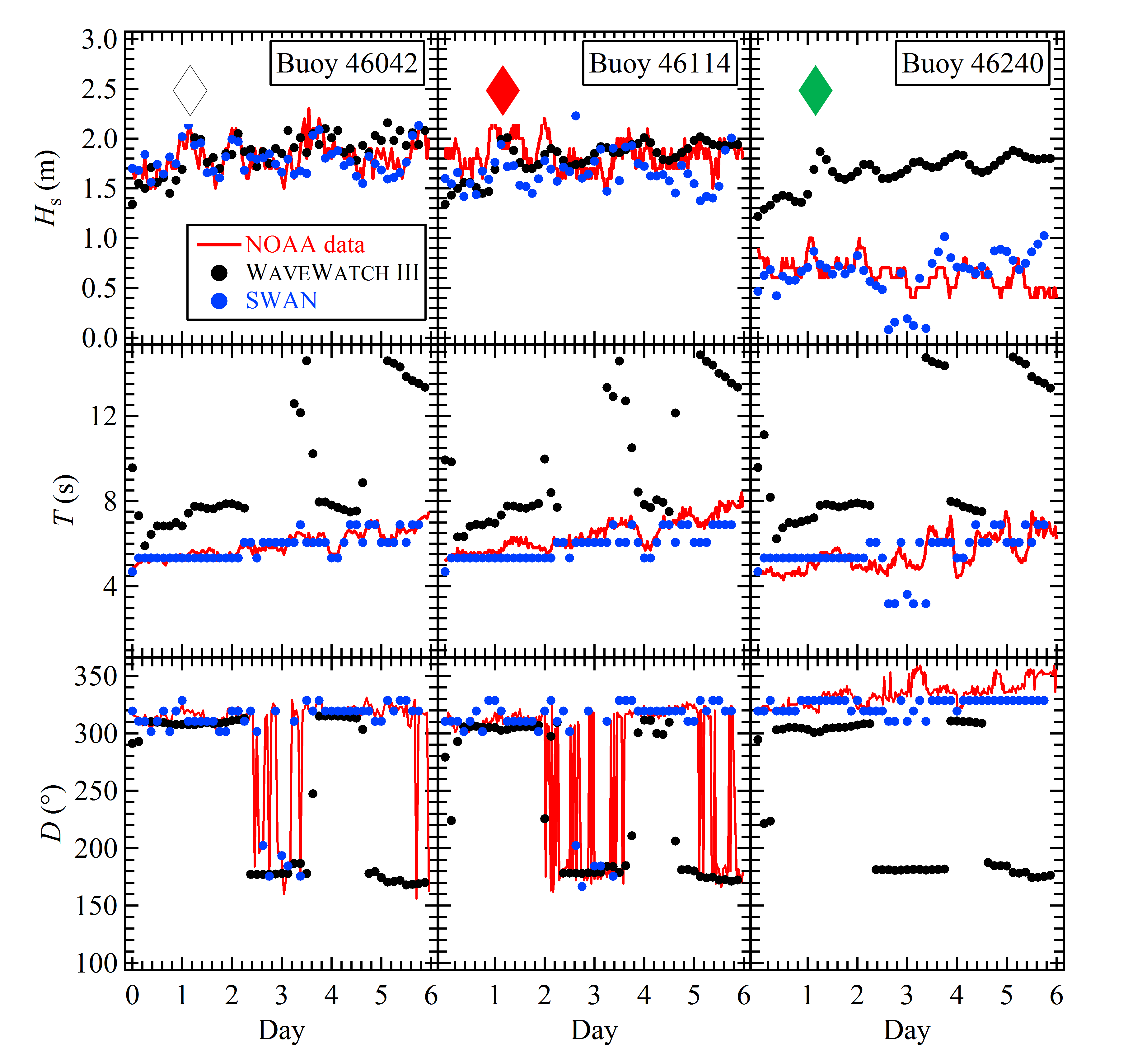}
\caption{Comparison of measured and simulated wave conditions at the three NOAA buoys.}
\label{fig:verification}
\end{figure}

Six days of NOAA wave data, ROMS ocean currents, and TWC winds were assembled into \emph{steady-state} SWAN model runs at three-hour intervals. Figure~\ref{fig:SWANHs} is an example SWAN-simulated $H_\mathrm{s}$ field showing waves entering the domain from the north-northwest and diffracting around the northern coast of Monterey Bay.  Figure~\ref{fig:verification} compares NOAA wave-condition data (significant wave height, $H_\mathrm{s}$, top row; wave period, $T$, middle row; wave direction, $D$, bottom row) for the three buoys in the Monterey Bay area (red curves) to W{\small AVE}W{\small ATCH}~III forecasts (black symbols) at that model's grid point nearest each buoy. Given the vast extent of the relatively coarse W{\small AVE}W{\small ATCH}~III model domain ($0.25^\circ$ resolution for the entire Eastern North Pacific compared to $0.001^\circ$ of the SWAN model), it is not surprising that there is a degree of local mismatch. Moreover, the location of NOAA Buoy~46240 (green diamond in Figure~\ref{fig:bathymetry}), which is sheltered from incoming westward waves, results in a notable discrepancy between W{\small AVE}W{\small ATCH}~III-simulated and measured wave conditions. This is expected because the nearest W{\small AVE}W{\small ATCH}~III model node is 15~km away at the turquoise circle to the northwest of the green diamond in Figure~\ref{fig:bathymetry}. Blue symbols are SWAN-simulated wave conditions when NOAA wave conditions from Buoy~46042 were supplied to the SWAN model as boundary conditions. SWAN does a good job of capturing the effects of bathymetry and coastline on the wave conditions at Buoy~46240 given that it is forced by the wave conditions displayed as the black curves (NOAA data) in the left column of Figure~\ref{fig:verification}. Overall, SWAN is able to simulate wave conditions more accurately than the W{\small AVE}W{\small ATCH}~III model. Root-mean-squared errors (RMSEs) are listed in Table~\ref{table:RMSE}. \citet{bidlot2002intercomparison} note that 40- to \mbox{60-cm} RMSEs are typical for $H_\mathrm{s}$. The simulations from the SWAN model forced by NOAA wave-condition data yield an appropriate (RMSE~$<60$~cm) match to available NOAA data indicating that the SWAN model is acceptable for developing the machine learning model training data set.

\begin{table}\footnotesize
\caption{RMSEs between NOAA wave-condition data and {\normalsize W}AVE{\normalsize W}ATCH~{\normalsize III} simulations at the grid point nearest to the indicated buoy and SWAN simulations when forced by data from Buoy~46042.}
    \label{table:RMSE}
    \begin{tabular}{| l | c | c | c | c | c | c | c | c | c |}
    \hline
    \multicolumn{1}{|c|}{Model} & \multicolumn{3}{|c|} {Buoy 46042} & \multicolumn{3}{|c|} {Buoy 46114} & \multicolumn{3}{|c|} {Buoy 46240}\hspace{-0.28cm} \\ \cline{2-10}
     &$H_\mathrm{s}$ (cm)&$T$ (s)&$D$ ($^\circ$)&$H_\mathrm{s}$ (cm)&$T$ (s)&$D$ ($^\circ$)&$H_\mathrm{s}$ (cm)&$T$ (s)&$D$ ($^\circ$) \\    \hline
    {\normalsize W}AVE{\normalsize W}ATCH~{\normalsize III}                 & 67 & 3.61 & 80.3 & 62 & 3.29 & 74.5 & 162 & 4.08 & 103.7 \\    \hline
    SWAN          & 50 & 0.57 & 62.3 & 41 & 0.45 & 73.9 & 42 & 1.06 & 12.0  \\
    \hline
\end{tabular}
\end{table}

For production runs, the SWAN model resolution was coarsened from $0.001^\circ$ to $0.01^\circ$; this yielding 3,104 active nodes in the domain. This model could be run in under 10~seconds as as compared with  the refined model, which took minutes to calculate and resulted in 224,163 active nodes. Although solving the refined SWAN model is only a matter of computational resource, the analysis of the spatial covariance indicates that $0.001^\circ$ resolution is not necessary to develop the framework outlined here; neither is a wave field comprising 224,163 $H_\mathrm{s}$ values.

\section{Machine Learning}
\label{S:2}

Two different supervised machine learning models are used to perform two different tasks: regression analysis for wave height and classification analysis for characteristic period. The multi-layer perceptron (MLP) conceptual model used to replicate $H_\mathrm{s}$ is loosely based on the anatomy of the brain. Such an artificial neural network is composed of densely interconnected information-processing nodes organized into layers. The connections between nodes are assigned ``weights,'' which determine how much a given node's output will contribute to the next node's computation. During training, where the network is presented with examples of the computation it is learning to perform (i.e.,~SWAN model runs), those weights are optimized until the output of the network's last layer consistently approximates the result of the training data set (in this case, wave heights). A~support vector machine (SVM) classification analysis constructs hyperplanes (planes in high-dimensional space) that divide the training data set into labeled groups (in this case, characteristic periods). 

\subsection{Background}
Machine learning has shown enormous potential for pattern recognition in large data sets. Consider that a physics-based model like SWAN acts  as a nonlinear function that transforms inputs (wave-characteristics boundary conditions and the spatially variable ocean currents and wind speeds) to outputs (spatially variable $H_\mathrm{s}$ and characteristic $T$). These data and corresponding simulations can be assembled into an input vector, $\underline{\mathbf{x}}$, and an output vector, $\underline{\mathbf{y}}$, respectively.
 
Because the goal of this effort is to develop a machine learning framework to act as a surrogate for the SWAN model, the nonlinear function mapping inputs to the best representation of outputs, $\hat{\underline{\mathbf{y}}}$, is sought:
\begin{equation}
\label{eqn:4}
	  	g\left({\underline{\mathbf x}};\underline{\underline{\mathbf{\Theta}}}\right)=\hat{\underline{\mathbf{y}}}.
\end{equation}

A~sufficiently trained machine learning model yields a mapping matrix, $\underline{\underline{\mathbf{\Theta}}}$, that can act as a surrogate for the SWAN model. This facilitates sidestepping of SWAN model by replacing the solution of the partial differential equation with the data-driven machine learning model composed of the vector-matrix operations encapsulated in (\ref{eqn:4}). 

Python toolkit SciKit-Learn \citep{pedregosa2011scikit} was used to access high-level programming interfaces to machine learning libraries and to cross validate results. Two distinct machine learning models were implemented here: an MLP model for $H_\mathrm{s}$ and an SVM model for characteristic $T$. Two different approaches were undertaken because the discrete characteristic $T$ values were more accurately represented by an SVM model than an MLP model (more on this later).

\subsubsection{The Multi-layer Perceptron Model}
\begin{figure}
\centering
\includegraphics[width=1.0\textwidth]{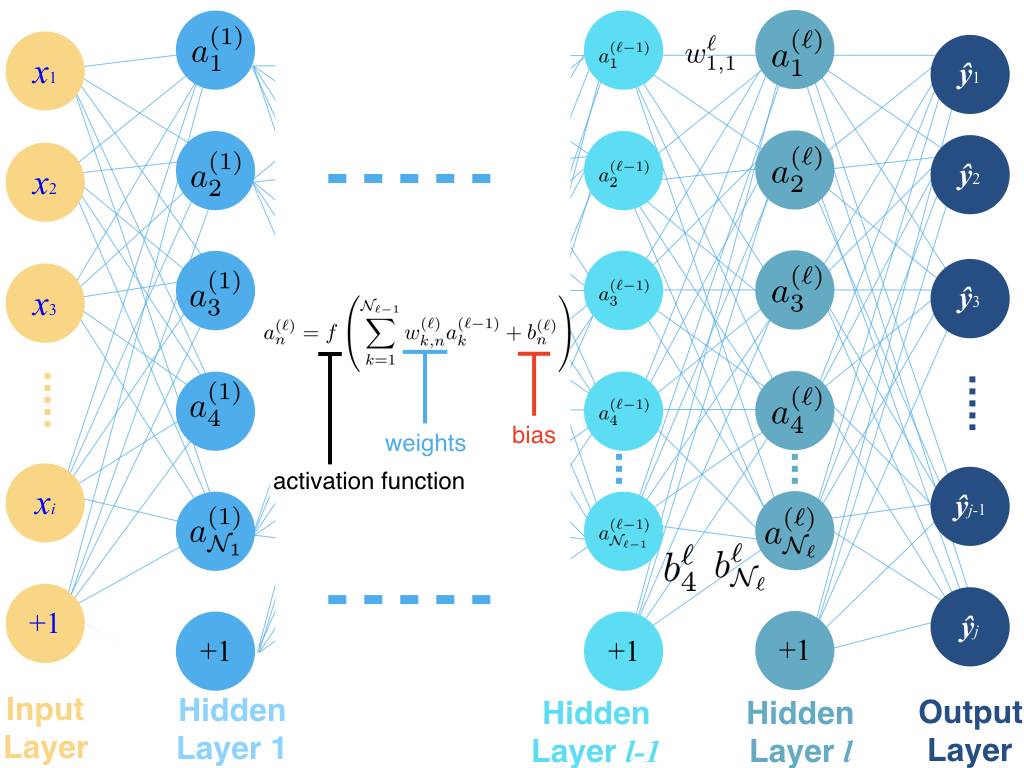}
\caption{Schematic of an MLP machine learning network.}
\label{fig:deeplearning}
\end{figure}

An MLP model is organized in sequential layers made up of interconnected neurons. As illustrated in Figure~\ref{fig:deeplearning}, the value of neuron $n$ in hidden layer $\ell$ is calculated as:
\begin{equation}
a_{n}^{\left(\ell\right)} = f\left(\sum^{\mathcal{N}_{\ell-1}}_{k=1} w_{k,n}^{\left(\ell\right)} a_{k}^{\left(\ell-1\right)} + b_{n}^{\left(\ell\right)}\right),
\end{equation}
where $f$ is the activation function, $\mathcal{N}_{\ell-1}$ is the number of nodes in layer $\ell-1$, $w_{k,n}^{\left(\ell\right)}$ is the weight projecting from node $k$ in layer $\ell-1$ to node $n$ in layer $\ell$, $a_{k}^{\left(\ell-1\right)}$ is the activation of neuron $k$ in hidden layer $\ell-1$, and $b_{n}^{\left(\ell\right)}$ is the bias added to hidden layer $\ell$ contributing to the subsequent layer.  The activation function selected for this application was the rectified linear unit (ReLU) \citep{vinod2010relu}:
\begin{equation}
f\left(z\right)=\max\left(0,z\right).
\end{equation}

A~loss function is defined in terms of the squared error between the SWAN predictions and the machine-learning equivalents plus a regularization contribution:
\begin{equation}
\vartheta=\frac{1}{2}\sum_{k=1}^m||{\underline{\mathbf y}}^{\left( k\right)}-{\hat{\underline{\mathbf{y}}}}^{\left( k\right)}||_2^2 + \alpha||\underline{\underline{\mathbf{\Theta}}}||_2^2,
\end{equation}
where the $||\cdot||_2$ indicates the $L_2$ norm. The regularization term  penalizes complex models by enforcing weight decay, which prevents the magnitude of the weight vector from growing too large because large weights can lead to overfitting where, euphemistically, the machine learning model ``hallucinates'' patterns in the data set  \citep{goodfellow2016deep}.

By minimizing the loss function, the supervised machine learning algorithm identifies the $\underline{\underline{\mathbf{\Theta}}}$ that yields $\underline{\mathbf{\hat{y}}}\approx\underline{\mathbf y}$. As shown in Figure~\ref{fig:deeplearning}, a machine learning model transforms an input vector (layer) to an output layer through a number of hidden layers. The machine learning model is trained on a data set to establish the weights parameterizing the space of nonlinear functions mapping from $\underline{\mathbf x}$ to $\underline{\mathbf y}$. Of course, a large training data set is required to develop a robust machine learning model; one luxury of the approach developed here is that such a data set is straightforward to assemble by completing many thousands of SWAN model runs to accumulate many input vectors, $\underline{\mathbf x}^{\left(m\right)}$, into an $\left(m\times i\right)$ design matrix, $\underline{\underline{\mathbf X}}$, which when acted upon by functions representing the hidden layers, that is, the $\left(i\times j\right)$ $\underline{\underline{\mathbf{\Theta}}}$ mapping matrix, yields the $\left(m\times j\right)$ output matrix $\underline{\underline{\mathbf Y}}$ whose rows comprise $\underline{\mathbf{\hat y}}^{\left(m\right)}\approx\underline{\mathbf{y}}^{\left(m\right)}$. Note that in practice, $\underline{\underline{\mathbf{\Theta}}}$ consists of a set of matrices. Specifically, for each layer there is a matrix, $\underline{\underline{\mathbf W}}$, of size   $\left(\mathcal{N}_{\ell-1}+1\right)\times\mathcal{N}_\ell$, comprising the optimized layer weights augmented with a column containing the biases for each layer. To determine the activation of the neurons in the next layer, its transpose is multiplied by the preceding layer's activation:
\begin{equation}
\underline{\mathbf a}^{\left(\ell\right)}=\left[{\underline{\underline{W}}}^{\left(\ell\right)}\right]^{\mathrm{T}}\cdot\underline{\mathbf a}^{\left(\ell-1\right)}+\underline{\mathbf b}^{\left(\ell\right)}
\end{equation}

The loss function was minimized across the training data set using an adaptive moment (Adam) estimation optimization method \citep{kingma2014adam}, which is a stochastic gradient-based optimizer. SciKit-Learn's default value of $\alpha=0.0001$ was used; a small $\alpha$ helps minimize bias. The MLP model finds weights and biases that minimize the loss function.

\subsubsection{Support Vector Machine Model}
A~One-versus-One (OvO) SVM multi-class classification strategy was applied to replicate the discrete $T$ values from the SWAN simulations \citep{Knerr1990ovo}. SVMs are binary classifiers but the OvO approach can handle multiple classes. OvO assembles all combinations of characteristic-$T$ pairs into $\mathsf{N}(\mathsf{N}-1)/2=55$ binary classes (because there were $\mathsf{N}=11$ discrete values for $T$ in the entire design matrix). The training data are divided into groups corresponding to the 55 combinations of $T_\mathsf{i}T_\mathsf{j}$ pairs (e.g.,~all $\underline{\mathbf x}$ vectors associated with $T_1$ and $T_2$, then all $\underline{\mathbf{x}}$ vectors associated with $T_1$ and $T_3$ all the way up to all $\underline{\mathbf{x}}$ vectors associated with $T_{\mathsf{N}-1}$ and $T_\mathsf{N}$). A~hinge-loss function (including $L_2$ regularization) is defined as \citep{moore2011l1}: 
\begin{equation}
\vartheta=\frac{1}{\mathsf n}\sum_{k=1}^{\mathsf n} \left\{\max\left[ 0,1-\psi\left({\underline{\mathbf w}}^{\mathrm{T}}\cdot{\underline{\mathbf x}}^{\left( k\right)}+b\right)\right]\right\}^2+\alpha||\underline{\mathbf w}||_2^2.
\end{equation}
where  $\psi=\pm 1$ distinguishes members of the $T_\mathsf{i}T_\mathsf{j}$ pair (i.e.,~$\psi=+1$ for $T_\mathsf{i}$ and $\psi=-1$ for $T_\mathsf{j}$) and $\mathsf{n}$ is the number of training-data vectors in the $T_\mathsf{i}T_\mathsf{j}$ group. Note that the hinge loss is 0 when $\psi\left(\underline{\mathbf w}^{\mathrm{T}}\cdot\underline{\mathbf x}+b\right) \geq 1$. The cost function is large when $T_\mathsf{i}$ is associated with $\underline{\mathbf w}^{\mathrm{T}}\cdot\underline{\mathbf x}+b\ll 0$ or when $T_\mathsf{j}$ is associated with $\underline{\mathbf w}^{\mathrm{T}}\cdot\underline{\mathbf x}+b\gg 0$. 
Minimization of the cost function results in selection of weights that avoids associating $T_\mathsf{j}$ with $\underline{\mathbf w}^{\mathrm{T}}\cdot\underline{\mathbf x}+b\gg 0$ or $T_\mathsf{i}$ with $\underline{\mathbf w}^{\mathrm{T}}\cdot\underline{\mathbf x}+b\ll 0$. A~unique $\underline{\mathbf w}$ is issued for each of the 55 training-data groups. These optimized $\left(\underline{\mathbf w},b\right)$ are aggregated into the $\left(i\times\mathsf{N}\right)$ mapping matrix, $\underline{\underline{\mathbf\Theta}}$. Next, the dot product of a training-data vector with the mapping matrix yields 55 rational numbers, which are translated into ``votes.'' When the rational number is positive for a $T_\mathsf{i}T_\mathsf{j}$ pair, a vote is cast for $T_\mathsf{i}$ (for which $\psi=+1$) while a negative rational number is a vote for $T_\mathsf{j}$ (for which $\psi=-1$). The $T$ with the most votes is nominated as the characteristics $T$ returned from the machine learning model.

\subsubsection{Training Data Sets}
Design matrices were developed by completing 11,078 SWAN model runs dating back to the archived extent of ROMS currents nowcasts (from April 1$^\mathrm{st}$, 2013 to June 30$^\mathrm{th}$, 2017). A~Python script downloaded the ROMS netcdf data and extracted $u$ and $v$ velocities from the nodes in its 3-km grid that are within the SWAN model domain (see the 357 black circles in Figure~\ref{fig:bathymetry}). Although ROMS ocean-current nowcasts are available every six hours, these data were linearly interpolated to generate ocean-current fields every three hours to increase the number of possible SWAN models and expand the training data set. Available NOAA wave-condition data from Buoy~46042 were downloaded at times corresponding to the three-hour increments of ROMS ocean-currents data. Finally, TWC historical wind speeds were downloaded at $0.25^\circ$ increments throughout the SWAN model domain (12 turquoise circles in Figure~\ref{fig:bathymetry}). There were 1,090 occasions when data from Buoy~46042 or ROMS currents were missing and no model was run for those times. A~MatLab script was written to develop input files including the primary SWAN input file where wave conditions ($H_\mathrm{s}$, $T$, and $D$) on the boundaries were specified, the spatially variable ROMS ocean-currents files (357 values each for $u$ and $v$), and the spatially variable TWC winds files (12 values each for easterly and northerly wind components). These data were assembled into $\mathbf x$ vectors comprising: the three wave-characteristic boundary conditions ($H_\mathrm{s}$, $T$, and $D$), 357$\times$2 ocean currents, and 12$\times$2 wind speeds. Overall, design matrix $\underline{\underline{\mathbf{X}}}$ has 11,078 rows and 741 columns. 

A~different $\underline{\underline{\mathbf{Y}}}$ is required for the MLP and OvO algorithms. For the MLP algorithm, $\underline{\underline{\mathbf{Y}}}$ is composed of the 11,078 SWAN model runs (rows), each of which comprises 3,104 wave heights (columns) defining the $H_\mathrm{s}$ field. For the OvO algorithm, only a $\underline{\mathbf y}$ vector of 11,078 characteristic $T$ values is supplied.

Note that in practice, data in the design matrices are pre-processed. Specifically, $\underline{\underline{\mathbf X}}$ undergoes a global normal transform (i.e.,~all constituent members are scaled so that their overall distribution is Gaussian with zero mean and unit variance). No pre-processing of MLP's $\underline{\underline{\mathbf Y}}$ is required, but OvO's $\underline{\mathbf y}$ is recast into labels 0 through $\mathsf N-1$ corresponding to characteristic $T_1$ through $T_\mathsf{N}$ where N = 11.

The $\underline{\underline{\mathbf{X}}}$ and $\underline{\underline{\mathbf{Y}}}$ data were always randomly shuffled into two groups to form the training-data set composed of 90\% of the 11,078 rows of data with the test-data set the remaining 10\%. Mapping matrix $\underline{\underline{\mathbf{\Theta}}}$ was calculated using the training data set and then applied to the test data set and the RMSE between test data vector, $\mathbf{\underline{y}}$, and its machine-learning representation, $\mathbf{\underline{\hat{y}}}$ was calculated.

For the MLP approach, training was performed many times to identify the number of hidden layers and the number of nodes per layer that yield the lowest overall RMSE. In practice, the MLP model offers two data files; the first describes the normal transform applied to $\underline{\mathbf x}$, the dot product of which is taken with the data included in the second file defining $\underline{\underline{\mathbf{\Theta}}}$.

The OvO algorithm need only be supplied with $\underline{\underline{\mathbf{X}}}$ and the column vector of characteristic $T$ values assembled as $\mathbf y$. The data were again split into two groups with 90\% of the $\mathbf{\underline{x}}$ vectors randomly assembled into the training data set with the rest reserved for testing. The OvO model returns three files; the first describes the normal transform applied to $\underline{\mathbf x}$, the dot product of which is taken with mapping matrix $\underline{\underline{\mathbf{\Theta}}}$ defined in the second file, and the third file defines how to label $\mathbf y$ and the same file is used for converting $\mathbf{\underline{\hat{y}}}$ back into the characteristic $T$.

\subsubsection{Significant Wave Heights}
\begin{figure}
\centering
\includegraphics[width=1.0\textwidth]{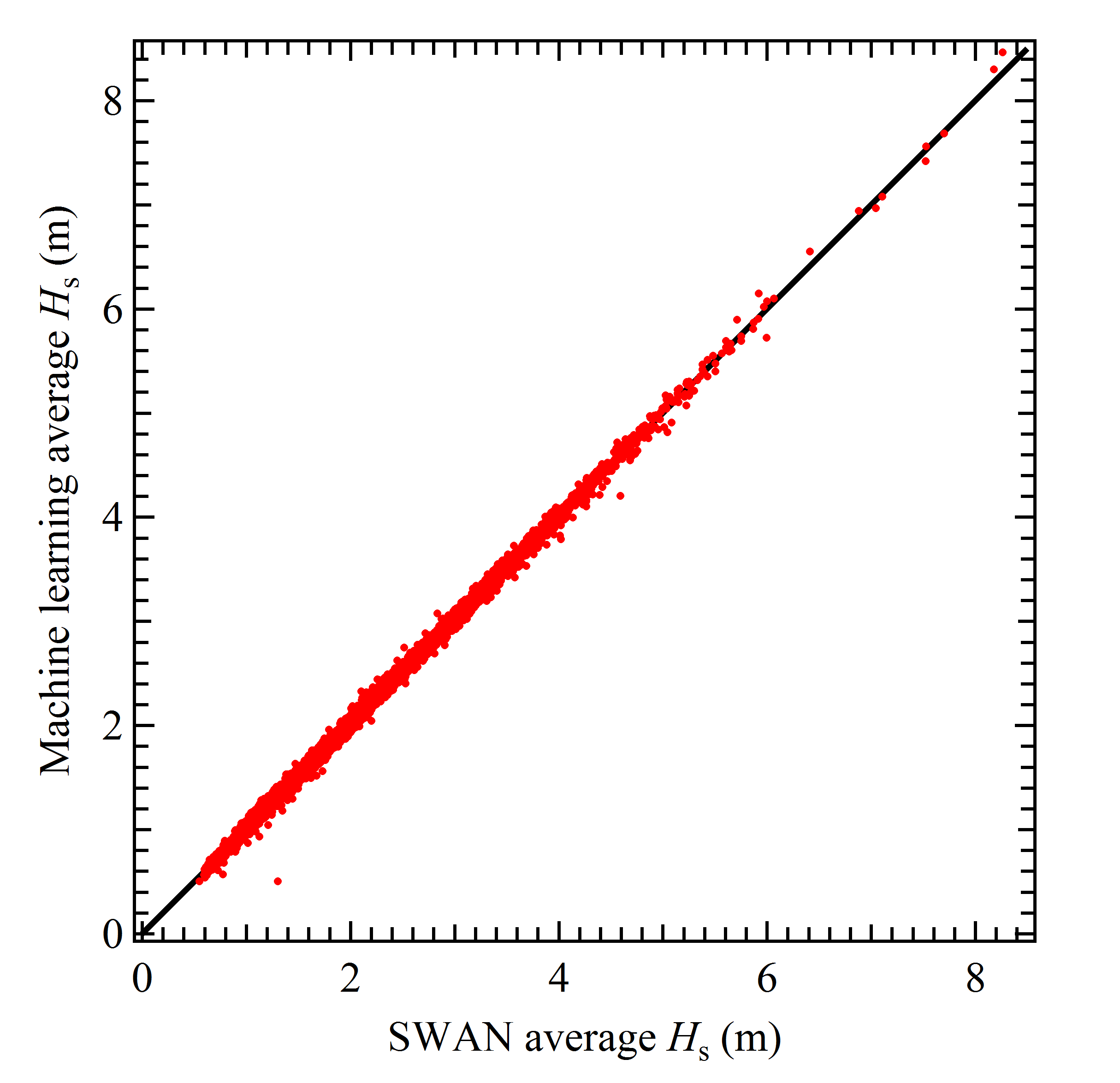}
\caption{Cross plot of SWAN and machine learning average $H_\mathrm{s}$ for each of the 11,078 model runs.}
\label{fig:CrossplotHs}
\end{figure}

MLP regression was used to reproduce the SWAN-generated $H_\mathrm{s}$. Initially, between two and 10 layers were investigated with anywhere from two to 3,000 nodes per layer, but it was quickly determined that fewer nodes (between 10 and 40 per layer) tended to yield smaller RMSEs for the test data used to evaluate each MLP layer/node combination. RMSEs ranged from 18~cm (six layers with 10 nodes each) to 9~cm (three layers with 20 nodes each), which is less than 5\% of the average $H_\mathrm{s}$. Although not appropriate for direct comparison, the RMSE for the MLP model is up to 80\% lower than those in the SWAN model with respect to the three buoy data sets (see Table~\ref{table:RMSE}). Figure~\ref{fig:CrossplotHs} summarizes the performance of the machine learning model at replicating SWAN-predicted wave heights showing the average $H_\mathrm{s}$ from each of the 11,078 SWAN model runs and the corresponding machine-learning estimates. A~line fit to the data in Figure~\ref{fig:CrossplotHs} has slope 1.002, so any bias is negligible. Moreover, note that even for the 14 instances where SWAN-simulated average $H_\mathrm{s}>6$~m, the machine learning representation was quite accurate (RMSE~$=14$~cm). In fact, the absolute relative error in the machine learning representation of wave height actually decreases with increasing average $H_\mathrm{s}$. That is, although the RMSE tends to increase with average $H_\mathrm{s}$, it does so at a slower rate than $H_\mathrm{s}$ itself. 

\begin{figure}
\centering
\includegraphics[width=0.5\textwidth]{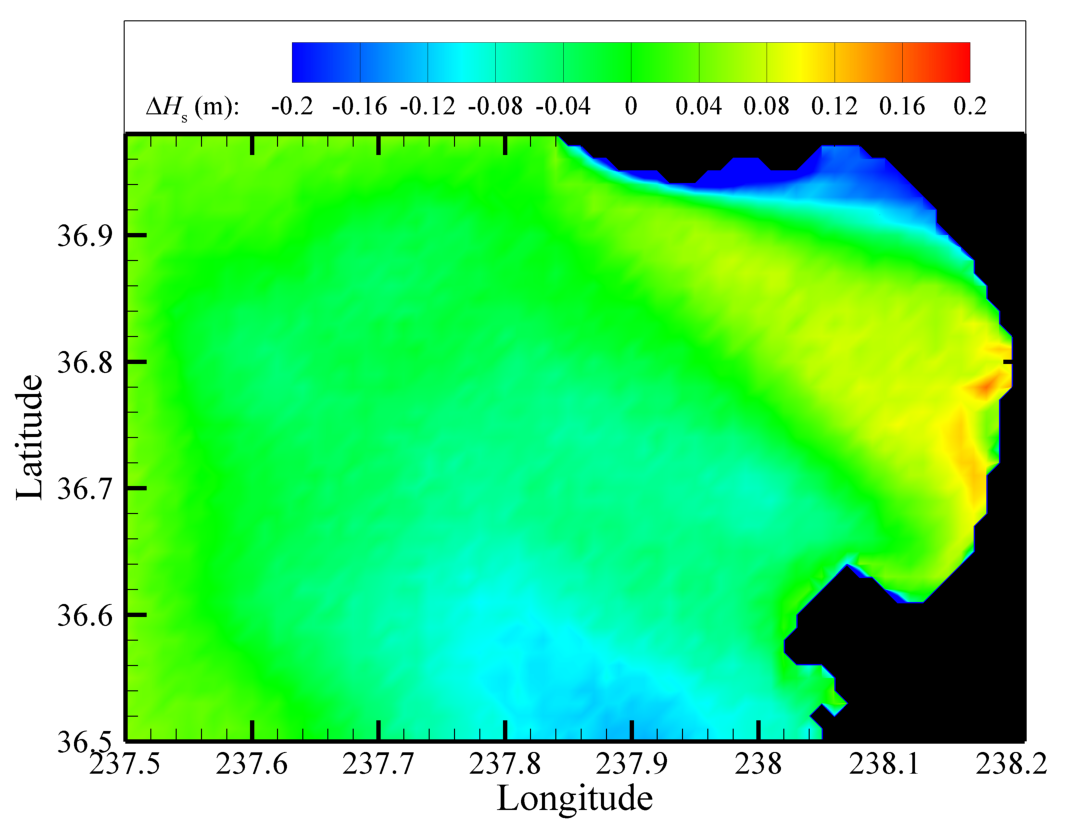}\includegraphics[width=0.5\textwidth]{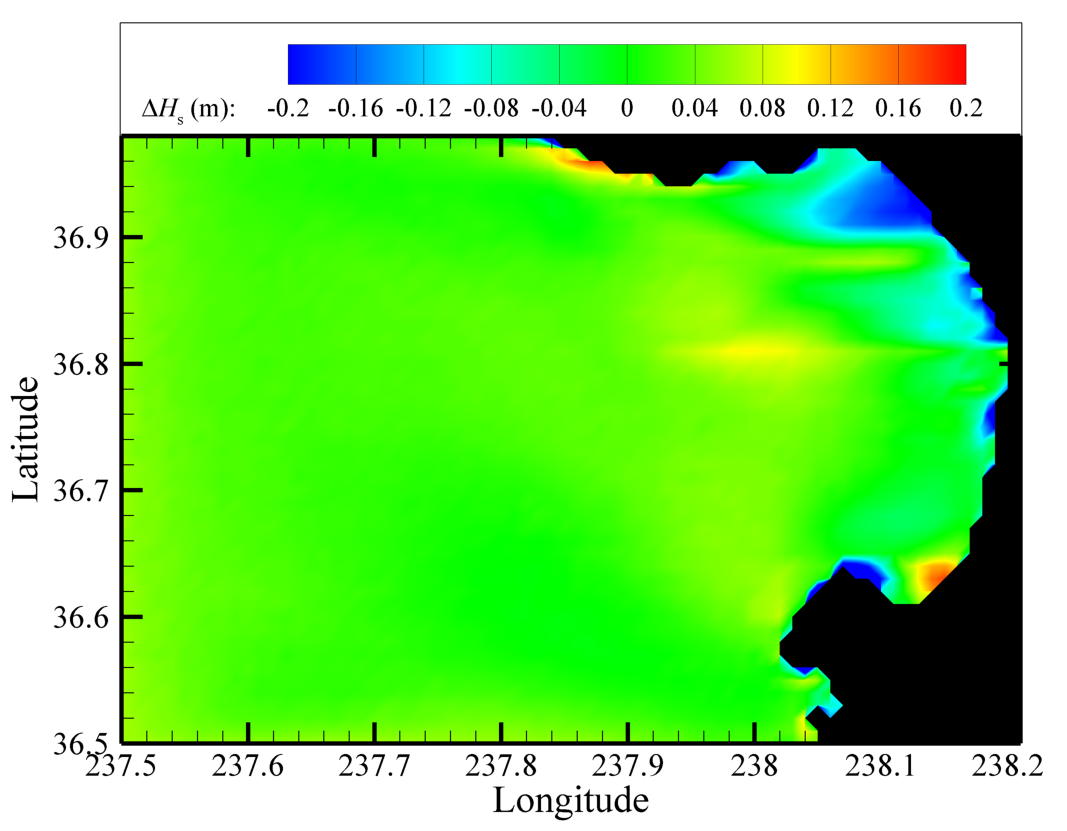}
\caption{Representative heat maps of the difference between SWAN- and machine-learning-simulated $H_\mathrm{s}$. The wave-height snapshot on the left shows some trends of local discrepancy (in this image, RMSE is 6~cm) not evident in the right figure, which actually has a higher RMSE (14~cm in this image).}
\label{fig:DeltaHs}
\end{figure}

Figure~\ref{fig:DeltaHs} shows representative contour plots  of the differences between SWAN-simulated $H_\mathrm{s}$ fields and the machine-learning equivalents. In the left snapshot, there remain some local trends where the machine learning model under-predicts $H_\mathrm{s}$ by up to 15~cm in the Bay and under-predicts $H_\mathrm{s}$ by 15~cm near the south boundary around longitude $237.9^\circ$ although RMSE~$=6$~cm. The right snapshot actually has a higher RMSE (14~cm), but does not reveal strong location-based trends. Many of the $H_\mathrm{s}$-differences snapshots were visualized and no clear trend in their patterns could be identified. (Perhaps this would benefit from another application of machine learning model like a convolutional neural network.)

A~$k$-fold cross validation \citep{bengio2004no} was conducted on the MLP model to assess whether overfitting has occurred and to ensure that the results can be generalized to an independent data set. In $k$-fold cross validation, the $\underline{\underline{\mathbf X}}$ input matrix is randomly partitioned into $k$ equal-sized subsamples. Of the $k$ subsamples, a single subsample is retained as the validation data for testing the model, and the remaining $k-1$ subsamples are used as training data. If some of the $k$-fold RMSes were notably higher than others, it could indicate over fitting or other model failings (z,~lucky selection of the test data set). Dividing the data set into 10 shuffled 90\%:10\%::train:test data sets (10~$k$-fold iterations) yielded RMSEs ranging from 8.0 to 10.2~cm for the test data set, which were always slightly outperformed, as expected, by the RMSEs for the training data set (ranging from 7.5 to 9.8~cm).

\subsubsection{Characteristic Wave Period}
\begin{figure}
\centering
\includegraphics[width=1.0\textwidth]{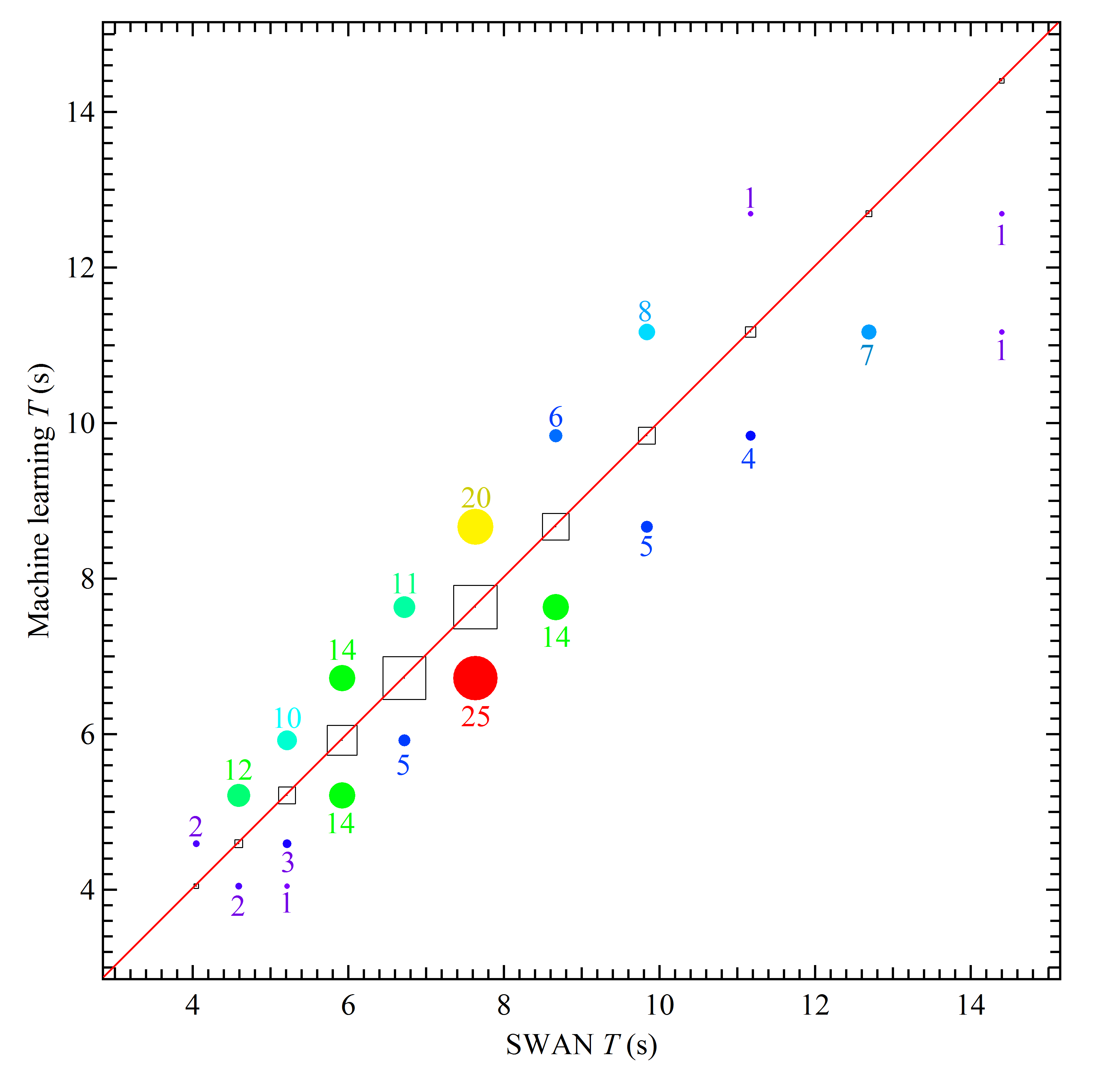}
\caption{Cross plot of SWAN and machine learning characteristic $T$. The number of missed $T$ values is indicated next to the symbol.}
\label{fig:CrossplotTper}
\end{figure}

Effectiveness of the OvO model was evaluated according to the percentage of correctly identified characteristic $T$ values. Unlike the MLP approach, no bias was observed in the OvO results and the percentage of characteristic $T$ accurately identified in the test data set was slightly higher than that from the best combination of layers and nodes in the MLP model (two layers of 10 nodes). OvO correctly identified the characteristic $T$ 90.1\% of the time in the test data set. The cross plot of characteristic $T$ from SWAN and the OvO  representation shown in Figure~\ref{fig:CrossplotTper} reveals that of the 166 times (out of 11,078 input vectors) that the machine learning model missed the characteristic $T$, it did so by one discretized $T$ increment except for two instances that were missed by two increments. Running 10 iterations of $k$-fold testing correctly identified the characteristic $T$ 90\% of the time in the  test data (and 98.6\% in the overall data set with an RMSE below 0.1~s).

It is worth noting that characteristic $T$ values were initially supplied to an MLP model after preprocessing $\bf\hat y$ into discrete values between 0 and $\mathsf{N}-1$. Again, various numbers of layers and nodes were used to replicate characteristic $T$ values and the percentage of accurate results in the test data was assessed. While the correct percentage of test data was comparable to the OvO scheme, there was a bias toward over-prediction; hence this approach was abandoned.

\section{Discussion}
Now that the mapping matrix (or vector) and the pre- and post-processor functions from the machine learning models have been developed, they can act as a surrogate for the SWAN model. Instead of historical data being used to build the input vector, $\bf x$, now forecast data can be used. To run in a forecast mode, the same model inputs used to force the SWAN must be assembled into a vector, which is multiplied by the machine learning mapping matrix to yield the $H_\mathrm{s}$ field and characteristic $T$. Such data are part of the Coastal Data Information Program \citep{o2016california} in the form of W{\small AVE}W{\small ATCH}~III-forecasted wave conditions available for up to the next 10~days. Also, ROMS-simulated ocean-currents and winds forecasts are available for the next 48~hours from the CeNCOOS \citep{CENCOOS} and \citet{TWC}, respectively. These forecast data are readily available for the Monterey Bay area as summarized in Table~\ref{table:online-data-descript}. In fact, an ensemble-based, machine learning framework is under development by the authors that provides improved wave-condition forecasts for Monterey Bay. 

\begin{table}\footnotesize
\caption{Forecast meta-data.}
\label{table:online-data-descript}
    \begin{tabular}{| c | c | c | c | c | }
    \hline
    Source & Data & Resolution & Times issued (UTC) & Forecast range (days)  \\ \hline \multicolumn{1}{|l|}
    {NOAA {\normalsize W}AVE{\normalsize W}ATCH~{\normalsize III}}   & $H_\mathrm{s}$, $T$, $D$ & $0.25^\circ$ & 0, 6, 12, 18 & 7.5 \\
    \hline \multicolumn{1}{|l|}
    {CeNCOOS ROMS}         & Ocean currents        & 3 km      & 3, 9, 15, 21 & 2  \\
    \hline \multicolumn{1}{|l|}
    {TWC}                  & E and N winds           & 1.1 km          & Hourly       & 14  \\
    \hline
\end{tabular}
\end{table}

The machine learning models can be executed to quickly generate the $H_\mathrm{s}$ field and corresponding characteristic $T$. Computationally, this only requires $L+1$ matrix multiplications. In fact, for a 48-hour forecast (16 simulations, one every every three hours), SWAN simulations on a single-core processor took 583~s (112~s on eight cores) while the machine-learning equivalent took 0.086~s to calculate the $H_\mathrm{s}$ field and 0.034~s to calculate the characteristic $T$ (a total of 0.12~s on a single processor) --- well over three orders of magnitude (485,833\%) faster than the running the full physics-based SWAN models. In fact, the operation requiring the most wall-clock time is actually loading the machine learning matrix files into memory. ``Edge computing'' \citep{shi2016edge} would have these mapping matrices pre-loaded into memory resulting in nearly instantaneous wave-field forecasts.

It is noted that the machine learning models presented here are specific to the Monterey Bay region and will require re-training to apply  to other locations. Of course, running a physics-based model at a new site requires grid generation and assembly of all boundary and forcing conditions and all of the attendant effort. However, what is important is that the framework necessary to develop this technology has been presented for the first time for wave modeling. It is expected that these sort of data-centric modeling machine learning approaches will grow increasingly common in the near future.

\section{Conclusions}Machine learning models have been developed as an accurate and computationally efficient surrogate for the SWAN model to respectively predict $H_\mathrm{s}$ fields and characteristic $T$. Using appropriately trained mapping matrices determined from supervised training of machine learning models, surrogates, which are really just matrix multiplication calculations, run over 4,000 times faster than the physics-based SWAN model and yield similarly accurate representations of wave conditions in the domain of interest. Thus, the machine learning models can act as a rapid, efficient wave-condition forecast system. These forecasted wave conditions can be used to estimate the power-generation potential of WECs or surf conditions. Ultimately, it is envisioned that such machine learning models could be installed locally on a WEC thereby facilitating it being its own forecast system. Moreover, the buoy itself can collect wave-condition data that can be used to update the machine learning models. As machine learning technologies improve, they can be adapted to compile a continuous stream of real-time data collected locally with available forecasts into ever-evolving and improving machine learning model parameters. In fact, such procedures are already frequently implemented with ``on-line learning'' \citep{cesa2004generalization}.

Additional efforts are currently underway to train a convolutional neural network (CNN) deep learning model to replicate the $H_\mathrm{s}$ field. Using an MLP approach does not allow for consideration of spatial information that could be contained in the data set. Specifically, augmenting the design matrix with additional data for the latitude, longitude, and bathymetric depth at each of the 3,104 SWAN model nodes will allow the CNN deep learning model to take into account how bathymetry affects wave heights and how, depending on incoming wave direction, waves are diffracted around the coastline.





\bibliographystyle{model1-num-names}
\bibliography{masterbib.bib}







\end{document}